\documentclass{emulateapj}
\usepackage{graphicx,verbatim}
\usepackage{amsmath}
\usepackage{amssymb}
\usepackage[usenames,dvipsnames,svgnames,table]{xcolor}

\usepackage{natbib}
\begin{document}
\def\etal{et al.\ \rm}
\def\ba{\begin{eqnarray}}
\def\ea{\end{eqnarray}}
\def\etal{et al.\ \rm}
\newcommand\cmtap[1]{{\color{blue}[SP: #1]}}
\newcommand\cmtas[1]{{\color{red}[AS: #1]}}
\title{Ab-Initio Pulsar Magnetosphere: Particle acceleration in Oblique Rotators and High-energy Emission Modeling}

\author{Alexander A. Philippov\altaffilmark{1}, Anatoly
  Spitkovsky}

\affil{Department of Astrophysical Sciences, 
Princeton University, Ivy Lane, Princeton, NJ 08540}
\altaffiltext{1}{sashaph@princeton.edu}

\begin{abstract}
We perform global particle-in-cell simulations of pulsar magnetospheres including pair production, ion extraction from the surface, frame dragging corrections, and high energy photon emission and propagation. In the case of oblique rotators, effects of general relativity increase the fraction of open field lines which support active pair discharge. We find that the plasma density and particle energy flux in the pulsar wind are highly non-uniform with latitude. Significant fraction of the outgoing particle energy flux is carried by energetic ions, which are extracted from the stellar surface. Their energies may extend up to a large fraction of the open field line voltage, making them interesting candidates for ultra-high-energy cosmic rays. We show that pulsar gamma-ray radiation is dominated by synchrotron emission, produced by particles that are energized by relativistic magnetic reconnection close to the Y-point and in the equatorial current sheet. In most cases, calculated light curves contain two strong peaks, in general agreement with Fermi observations. The radiative efficiency decreases with increasing pulsar inclination and increasing efficiency of pair production in the current sheet, explaining the observed scatter in $L_{\gamma}$ vs $\dot{E}$. We find that the high-frequency cutoff in the spectra is regulated by the pair loading of the current sheet. Our findings lay the foundation for quantitative interpretation of Fermi observations of gamma-ray pulsars.

\end{abstract}


\section{Introduction}

In recent years, significant progress in global particle-in-cell (PIC) modeling of pulsar magnetospheres has been made by several groups. It was shown how the magnetospheric structure changes from the charge-separated disk-dome configuration to the plasma-filled state, becoming close to the force-free solution under sufficient plasma supply from pair discharges \citep{Philippov14,Chen14,Belyaev15,Cerutti15,Philippov15a}. The effects of general relativity (GR) were shown to play an important role in establishing the pair production in aligned pulsars \citep{Philippov15b,Gralla16,Belyaev16}. Preliminary studies of nearly force-free magnetospheres showed the important role of the current sheet beyond the light cylinder for particle acceleration and emission of high-energy photons \citep{Cerutti16}. However, the pair supply in the magnetosphere was not modelled self-consistently in this work, and the question about identifying the emission region, as well as the properties of high-energy emission, remained.

In this paper we present GR PIC simulations of oblique rotators, including pair production by magnetic conversion of photons and by photon-photon collisions. These improvements allow for a self-consistent study of plasma production near the star and in the equatorial current sheet, as well as formation of light curves and spectra of high-energy radiation. The paper is organized as follows. In \S{\ref{sec:numerical-models}} we present improvements of our numerical method for pair production and radiation tracking. In \S{\ref{sec:psr}} we discuss the magnetospheric structure, mechanisms of particle acceleration and implications for the structure of pulsar wind. We present the modeling of pulsar high-energy emission in \S{\ref{sec:highenergy}}. We conclude and outline the prospects for future work in \S{\ref{sec:discussion}}.

\section{Numerical method}
\label{sec:numerical-models}
Our simulations were performed using 3D relativistic PIC code TRISTAN-MP. Our basic setup follows \citet{Philippov15a} (hereafter, PSC15) with a few modifications. Inside the star, we force the fields to known values with a smoothing kernel. The magnetic field in the star is set to the field of a rotating dipole, ${\vec{B}}=(3{\vec r}({\vec{\mu}}\cdot{\vec r})-{\vec \mu})/r^3$, where $\vec{\mu}(t)=(\cos({\Omega_*}t)\sin\alpha,\sin({\Omega_*}t)\sin\alpha,\cos\alpha)$ is the magnetic moment vector, $\Omega_*$ is the stellar angular velocity, and $\alpha$ is the inclination angle. Electric field inside the sphere is forced to corotation values $\vec{E}=-(\vec{\Omega}_{*}-\vec{\omega}_{LT}) \times \vec{r} \times \vec{B}/c$, where $\omega_{LT}$ is the Lense-Thirring angular velocity (see below). Between 2 and  4 $R_{LC}$ from the star we create the absorbing layer region \citep{Cerutti15}, which mimics the outflow boundary condition for both fields and particles. 

\citet{Philippov15b} showed that the effects of frame-dragging are essential for efficient pair production over a significant area of the polar cap. In order to take these into account, we modified the Maxwell solver of TRISTAN-MP and added a new term into Faraday's induction equation, which describes the generation of electric field due to the rotation of space-time:
\begin{equation}
\frac{1}{c}\frac {\partial{\vec B}} {\partial t}=-\nabla \times \left(\vec{E} +\frac{\vec{\beta}}{c}\times \vec{B}\right), \label{faraday eq}\\
\end{equation}
where $\vec{\beta} = \frac{1}{c} \vec{\omega}_{LT}\times \vec{r}$ and 
$\omega_{LT} = (2/5) \Omega_* (r_g/R_*) (R_{*}/r)^3$ (for details, see \citealp{Philippov15b}). Since frame-dragging is important only close to the stellar surface, the magnetic field in the RHS of (\ref{faraday eq}) can be considered to be dipolar. As was shown in \citet{Philippov15b}, \citet{Belyaev16}, and \citet{ Gralla16}, this is the only term which needs to be taken into account. All our simulations are performed for the compactness value of $r_g/R_*=0.5$. 

We consider two sources of plasma supply in the magnetosphere: extraction of particles from the stellar surface based on the value of local surface charge, similar to PSC15, and pair creation. Hereafter, positively charged particles coming from the surface are referred to as ions. Inside the light cylinder ions move  along the magnetic field lines, so their formal Larmor radii are not important for the dynamics of the magnetosphere. In the current sheet beyond the light cylinder magnetic reconnection proceeds in the relativistic regime, and the charge to mass ratio of each species is not very important \citep{Werner16}. This is why, for simplicity, we consider ions to have the same charge and mass as positrons; however, ions are not allowed to participate in the pair production process and do not experience radiation reaction force. In this paper, we consider two sources of pair production: magnetic conversion and the two-photon decay. The first channel operates near the star, at $r<2R_*$. We model the magnetic conversion in a way described in PSC15: a new pair is injected whenever the energy of a simulation particle exceeds  the threshold value, $\gamma > \gamma_{min}$, which is set to $\gamma_{min}=0.01\gamma_{0}=20$, where $\gamma_{0}$ is the Lorentz factor of a particle experiencing the full vacuum potential drop between the pole and the equator, $\gamma_0=(\Omega_* R_{*}/c)(e B_0 R_{*}/m_e c^2)\approx 2000$ in simulation. The polar cap voltage, $\Phi_{PC}$, corresponds to possible particle acceleration up to Lorentz factors $\gamma_{PC}=(\Omega_*R_*/c)\gamma_0=500$. The second channel mimics collisions of high-energy photons. Photons are emitted by energetic particles via synchro-curvature mechanism. Their frequency is set as $eB_{eff}\gamma^2/m_e c = e\sqrt{({\vec{E}}+{\vec{v}}\times{\vec{B}}/c)^2-({\vec{v}}\cdot {\vec{ E}}/c)^2}\gamma^2/m_e c$ \citep{Cerutti16}, where $\vec{v}$ and $\gamma$ are the velocity and Lorentz factor of the emitting particle. Momentum of the emitted photon is set to be along the direction of motion of the energetic particle. The emitted photons have energies much smaller than the particle energy, and hence the secondary pairs are created with Lorentz factors $\gamma_s \ll \gamma_{min}$. Energy loss due to the emission of synchro-curvature photons is explicitly taken into account in the particle equation of motion. Photons are tracked on the grid as computational particles. For each photon, we assign the mean free path, which scales with the local density of photons $n_{\gamma}^{-1}$, such that the typical mean free path of photons at the light cylinder is $L_{mfp}=0.5 R_{LC}$\footnote{The mean free path is calculated based on the value of local density at the photon emission point.}. The dependence of the mean free path on the local photon density mimics the two-photon collisions in the situation when the second photon is provided by the emission region itself, not by the external background (see Section 4 for details). We performed simulations with different values of $L_{mfp}$ and report our results below.

For a Crab-like pulsar $\gamma_{PC}\sim 10^{10}$ and $\gamma_{min} \sim 10^6$. The typical Lorentz factor of secondary pairs produced in $e^{\pm}$ discharge near the stellar surface is $\gamma_s \sim 10^2$. In our simulation we scale down these numbers, but preserve the hierarchy $\gamma_s \ll \gamma_{min} \ll \gamma_{PC}$. This ensures efficient pair production in the magnetosphere. Our simulation grid size is $2048^3$, the stellar radius $R_*$ is resolved with 120 computational cells, and $R_{LC}/R_*=4$. To study the wind structure at larger distances we performed simulations with half the resolution per stellar radius, which allowed to fit $8 R_{LC}$ in the simulation box. The plasma skin depth $\lambda_p$ is resolved with a few computational cells near the star, so that the hierarchy of spatial scales of typical pulsars, $R_{LC} > R_* \gg \lambda_p$, is also preserved\footnote{Typical pulsars have $R_*/\lambda_p\sim 10^6$ and $R_{LC}/R_* > 10^2$, while millisecond pulsars have $R_{LC}/R_* \sim 10$.}. To characterize the magnetic field strength we will use the magnetization parameter, $\sigma=B^2/(4 \pi n m_e c^2)$, where $B$ and $n$ are the local magnetic field strength and plasma density values, respectively. The plasma density is expressed in units of characteristic values of Goldreich-Julian density, so that the multiplicity parameter is defined as $\lambda=2\pi c n e/\Omega_* B$.

\section{Pulsar Magnetosphere}
\label{sec:psr}
\subsection{Magnetospheric structure}
\label{sec:magnetosphere}
\begin{figure*}
\centering
\includegraphics[width=\textwidth]{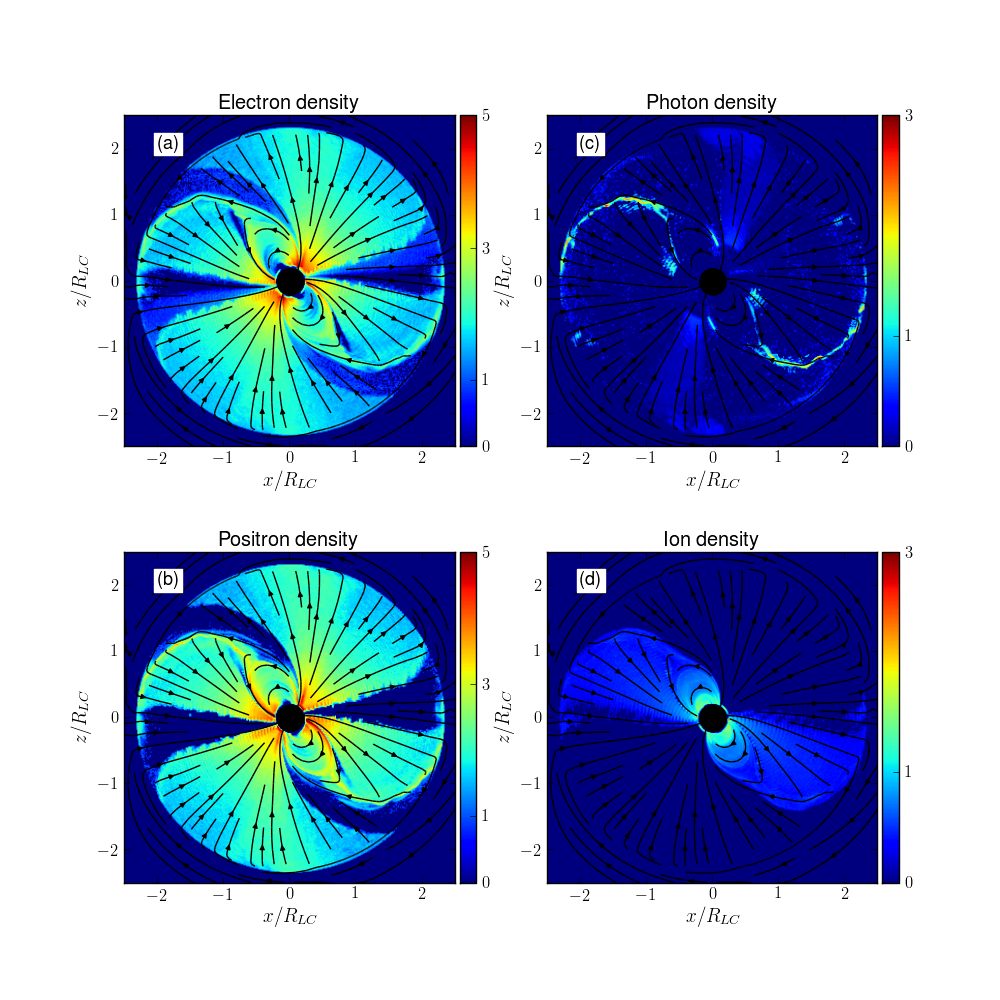}
\caption{Slice through the ${\vec \mu}-{\vec \Omega_*}$ plane for the $60^{\circ}$ inclined pulsar magnetosphere. (a): Electron density, (b): positron density, (c): photon density, (d): ion density. All densities are normalized by $\Omega_*B/2\pi ec$. Black lines show poloidal magnetic field lines.}
\label{fig:densities}
\end{figure*}

\begin{figure}
\centering
\includegraphics[width=0.5\textwidth]{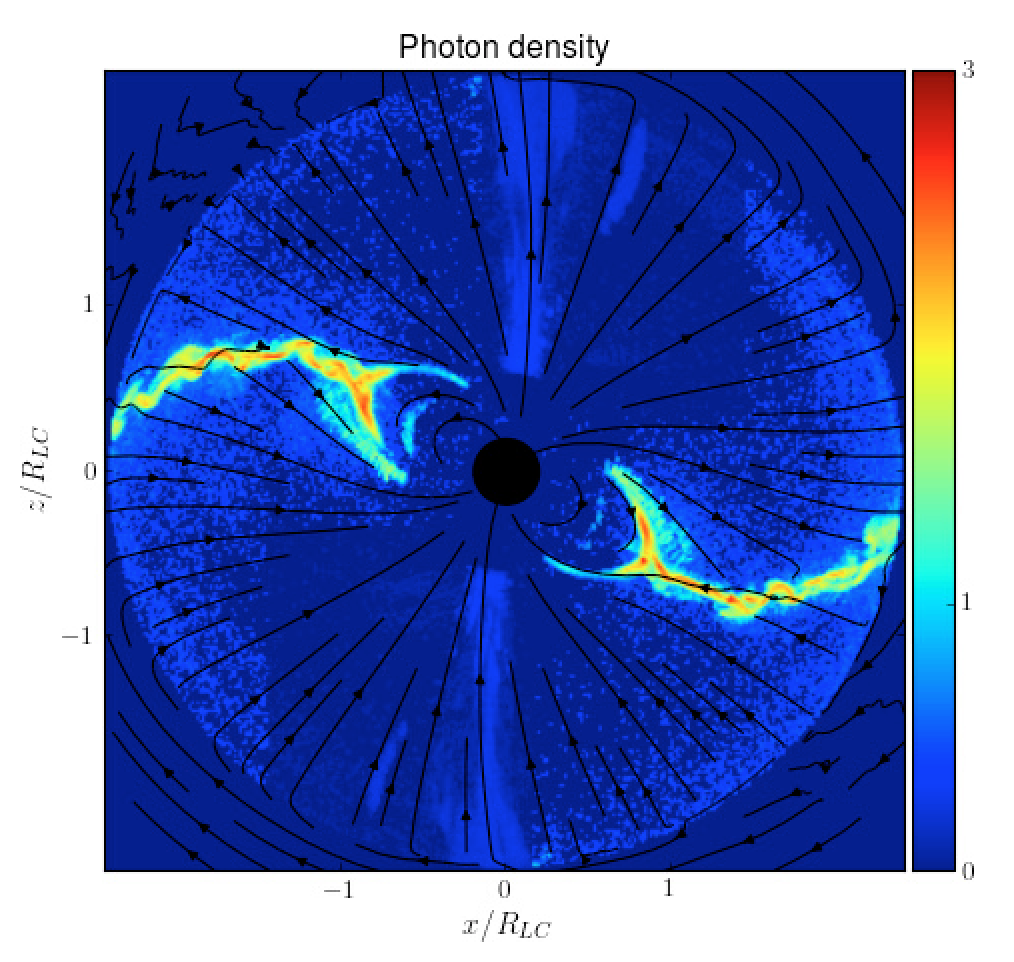}
\caption{High-energy photon density in the magnetosphere of $30^{\circ}$ rotator. Compared to the $60^{\circ}$ inclined solution, the return current layer inside the light cylinder is a significantly more efficient photon emitter. The kink instability of the current sheet beyond the light cylinder is also more active.}
\label{fig:30ph}
\end{figure}

After one rotation period, the solution approaches a steady-state in the corotating frame. We show the field structure and densities of electrons, positrons, high-energy photons, and ions in a slice through the 3D magnetosphere of $60^{\circ}$ rotator in Figure 1. The magnetic field structure and the spin-down luminosity of our pair-producing solutions are similar to the force-free solution and the flat space-time PIC simulation presented in PSC15. Compared to our earlier simulations in flat space-time, we find that the inclusion of GR effects significantly increases the number of pair producing field lines on the polar cap (compare Figure 1a,b here to Figure 3a,b in PSC15). However, as in the case of the aligned rotator, some field lines still carry sub-GJ current, which is supported by the charge-separated outflow of electrons \citep{Timokhin13,Philippov15b,Gralla16}. In the case of $60^{\circ}$ rotator, these field lines are located around the null current line $j_{\parallel}=0$ (see \S\ref{sect:acceleration} for details of polar discharge operation). In addition to the polar cap, we observe efficient particle acceleration and pair production in the return current layer and in the equatorial current sheet, which are highlighted by the distribution of high-energy photons emitted by energetic particles in the magnetosphere (Figure \ref{fig:densities}c). The voltage drop in the return layer is non-uniform with distance from the stellar surface, so that particle acceleration and, thus, pair production mostly takes place near the stellar surface and close to the Y-point. We note that for lower inclinations the return layer is a more efficient photon emitter (for example, see photon density in the magnetosphere of $30^{\circ}$ inclined pulsar in Figure \ref{fig:30ph}). However, for all inclinations, the current sheet beyond the Y-point is the most significant source of high-energy photons and pairs in the outer magnetosphere. Plasma in the closed field line region is mostly charge-separated and dominated by ions (see Figure \ref{fig:densities}d). These ions are extracted from the star with small momentum and travel along the dipolar field line until they hit the star in the other hemisphere. Ions are also pulled off the surface in  the return current regions and are accelerated to high energies in the magnetospheric current sheet, as discussed below. In the case of a significant mean free path of photons, we observe pair plasmas on some of the closed field lines. These pairs are produced by photons which are emitted close to the Y-point and propagate back towards the star. Pairs in the closed zone travel along the field lines until they hit the stellar surface. 

\begin{figure}
\centering
\includegraphics[trim=180px 55px 170px 55px, clip=true,width=0.5\textwidth]{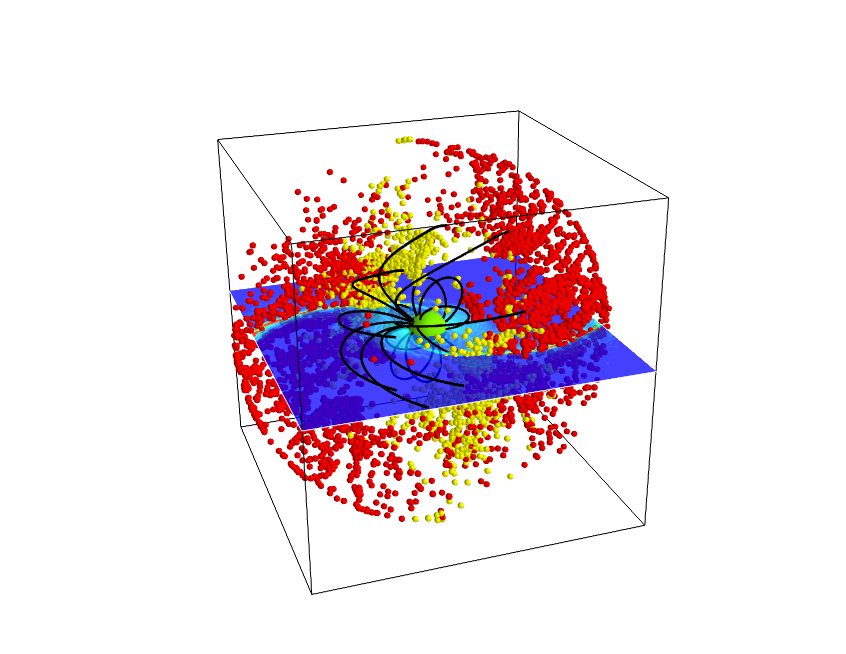}
\caption{Downsampled distribution of high-energy photons in the magnetosphere of $60^{\circ}$ inclined pulsar. Red dots show photons propagating outward, and yellow dots show photons which propogate back to the star. Region near the Y-point contains both kinds of photons. Color in the plane shows the mangetospheric current density.}
\label{fig:photons30}
\end{figure}

\begin{figure*}
\begin{center}
\includegraphics[width=0.49\textwidth]{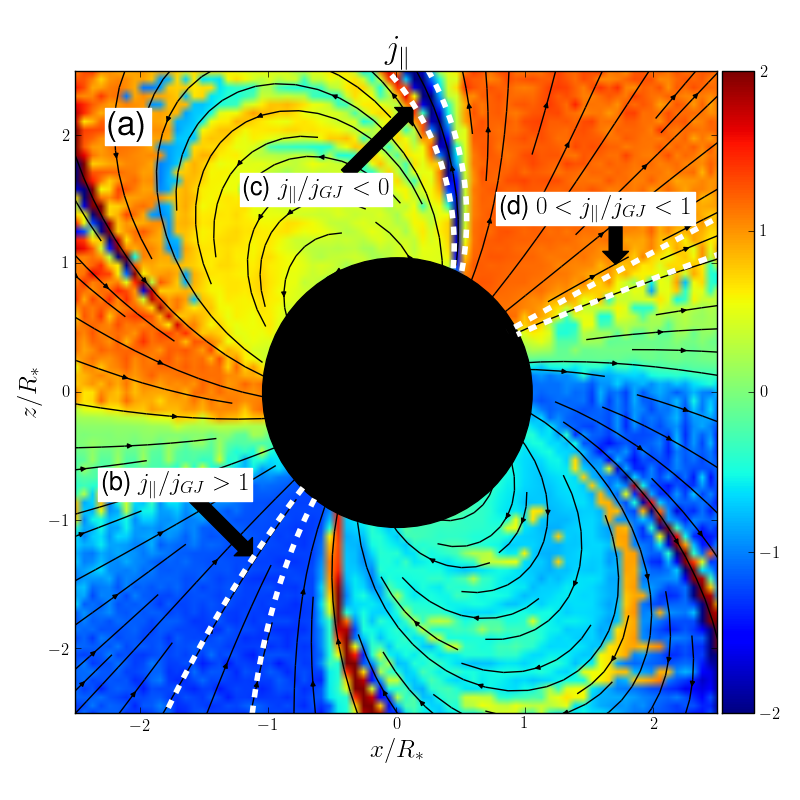} 
\hspace{\stretch{1}}
\includegraphics[width=0.49\textwidth]{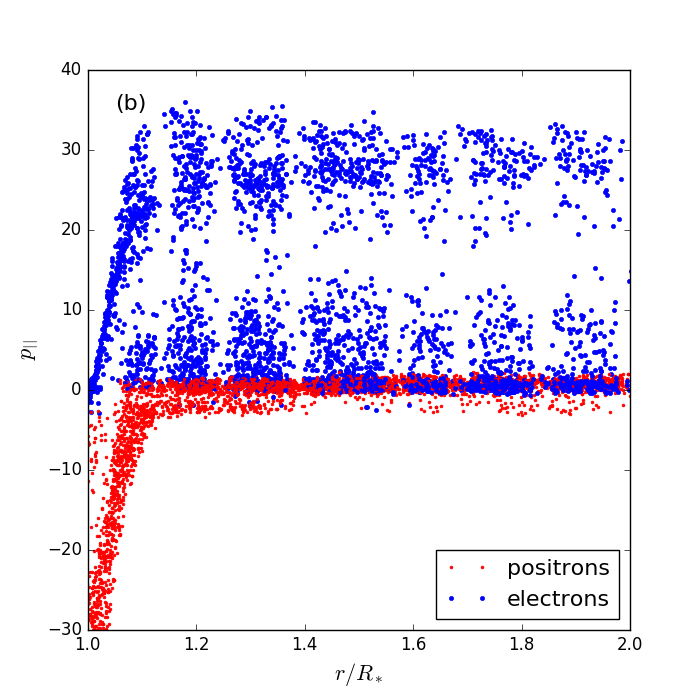} \\
\includegraphics[width=0.49\textwidth]{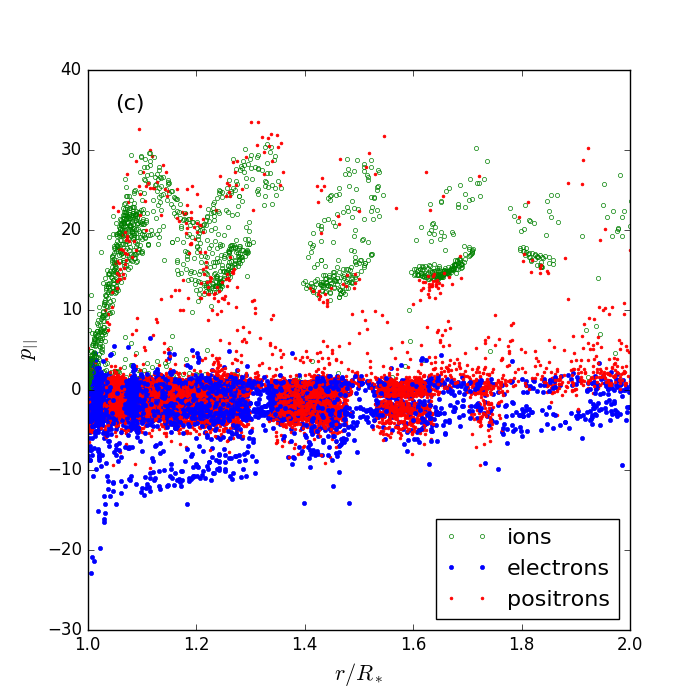} 
\hspace{\stretch{1}}
\includegraphics[width=0.49\textwidth]{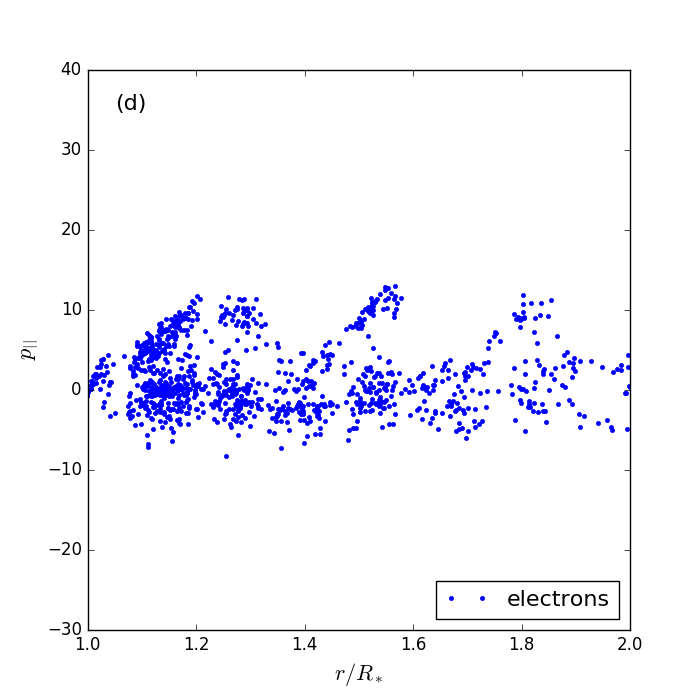}\\
\end{center}
\caption{Operation of polar discharge in the magnetosphere of $60^{\circ}$ inclined pulsar. (a) Current distribution in a slice through the ${\vec \mu}-{\vec \Omega_*}$ plane, marking three qualitatively different regimes of polar flow: (b) outgoing super-GJ current, $j_{\parallel}/j_{GJ}>1$; (c) return current; (d) outgoing sub-GJ current, $0<j_{\parallel}/j_{GJ}<1$. Black solid lines correspond to poloidal magnetic field lines, and color shows the current component parallel to the magnetic field, normalized by $\Omega_*B/2\pi$. Panels (b)-(d) show downsampled particle phase space distribution in three zones, bounded by white dashed field lines from panel (a). The dots of green, blue and red colors represent ions, electrons and positrons, respectively. Particle's momentum is normalized by $m_e c$ and the distance from the star $r$ is measured along the magnetic field lines and normalized by $R_*$.}
\label{fig:discharge}
\end{figure*}

\begin{figure}
\begin{tabular}{c}
  \includegraphics[width=75mm]{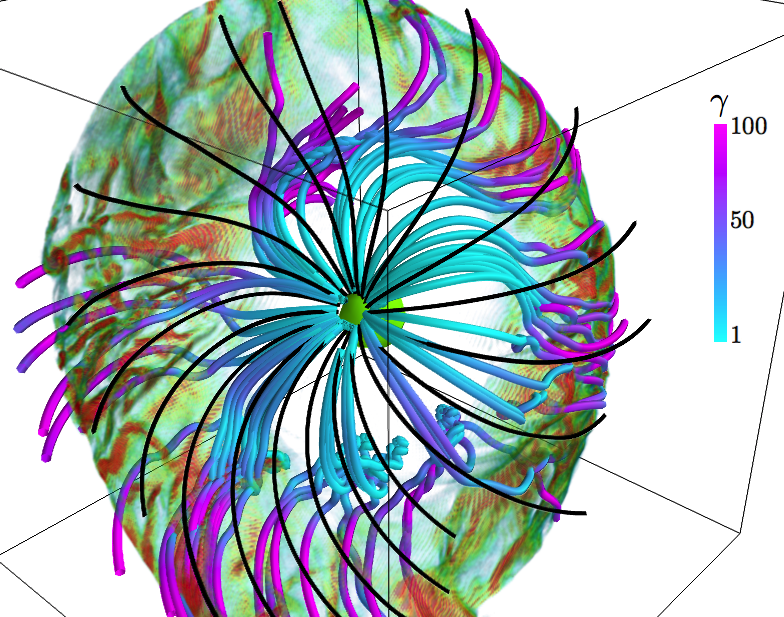} \\
  (a) Ions \\
  \includegraphics[width=75mm]{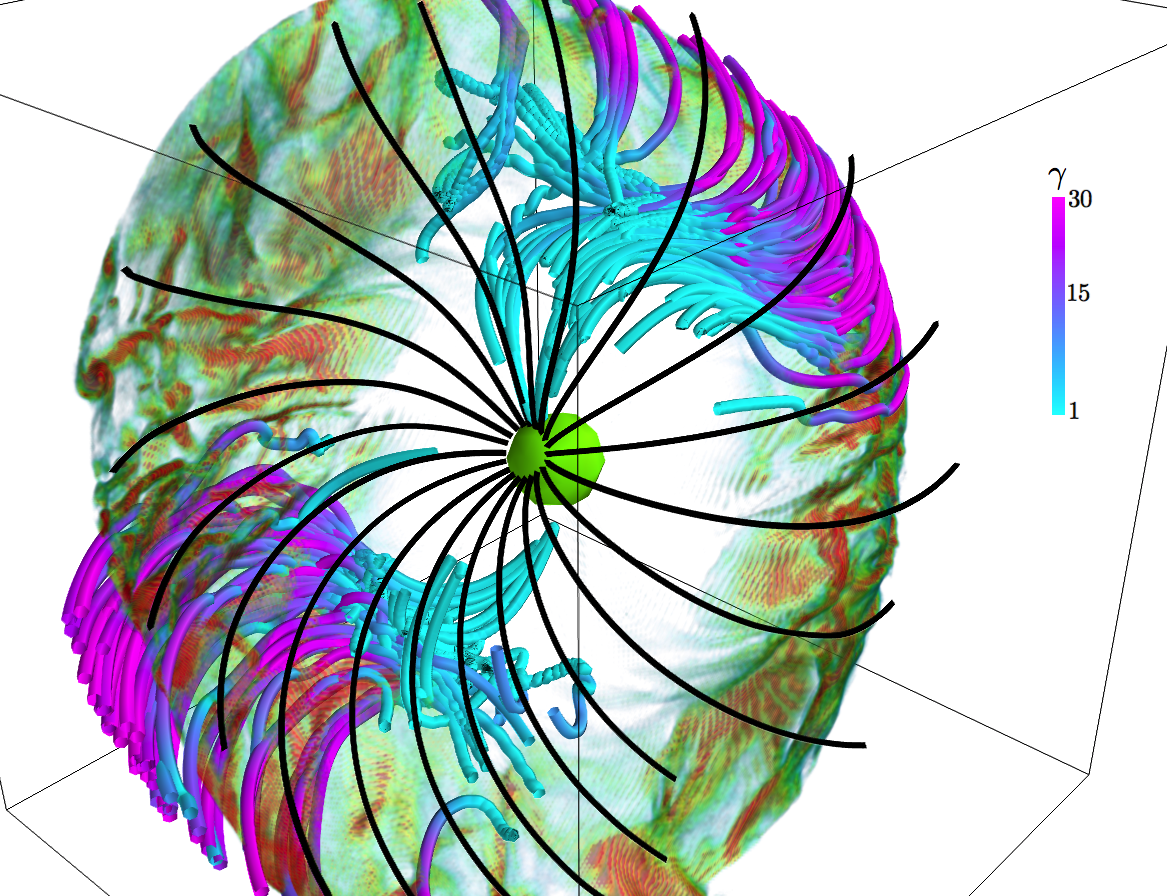} \\
  (b) Positrons \\
  \includegraphics[width=75mm]{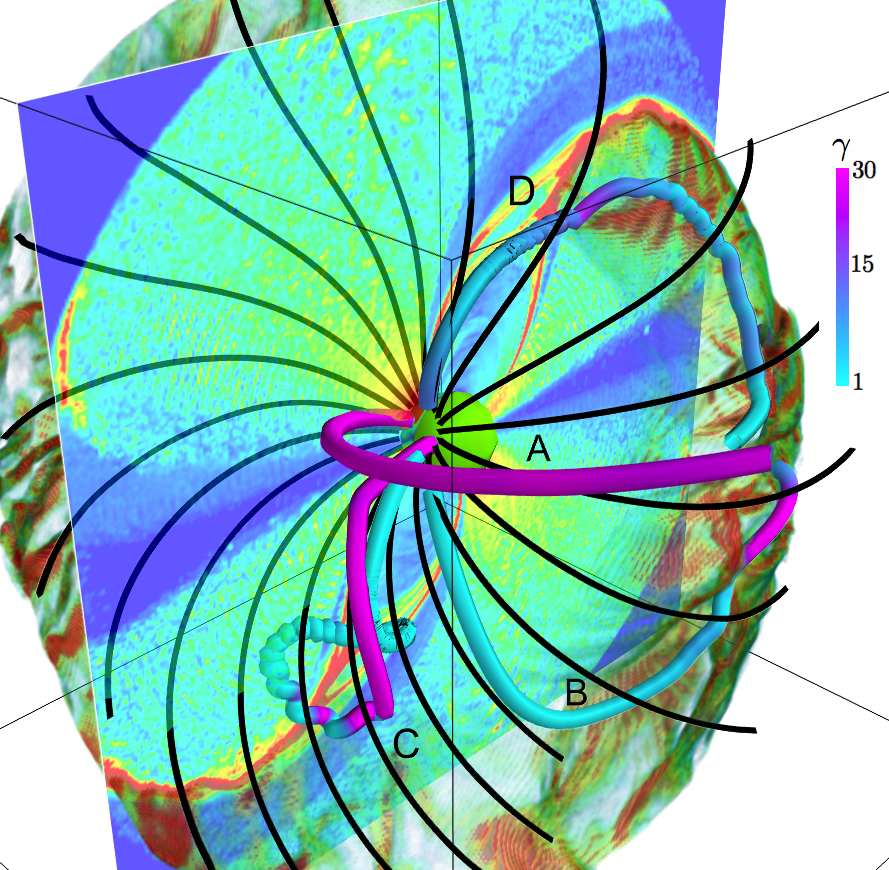}\\
  (c) Electrons
\end{tabular}
\caption{3D trajectories of particles in the corotating frame. Volume rendering shows current density, black lines show magnetic field lines, and colored thick lines show particle trajectories: (a) Ions. (b) Positrons. (c) Electrons. The color of trajectories shows particle Lorentz factor, with magenta color representing highest energies. Fewer electrons are shown to highlight different types of their trajectories. Color in plane in panel (c) shows the magnetospheric current density.}
\label{fig:particles}
\end{figure}

\begin{figure*}
\begin{tabular}{cc}
     \includegraphics[scale=0.4]{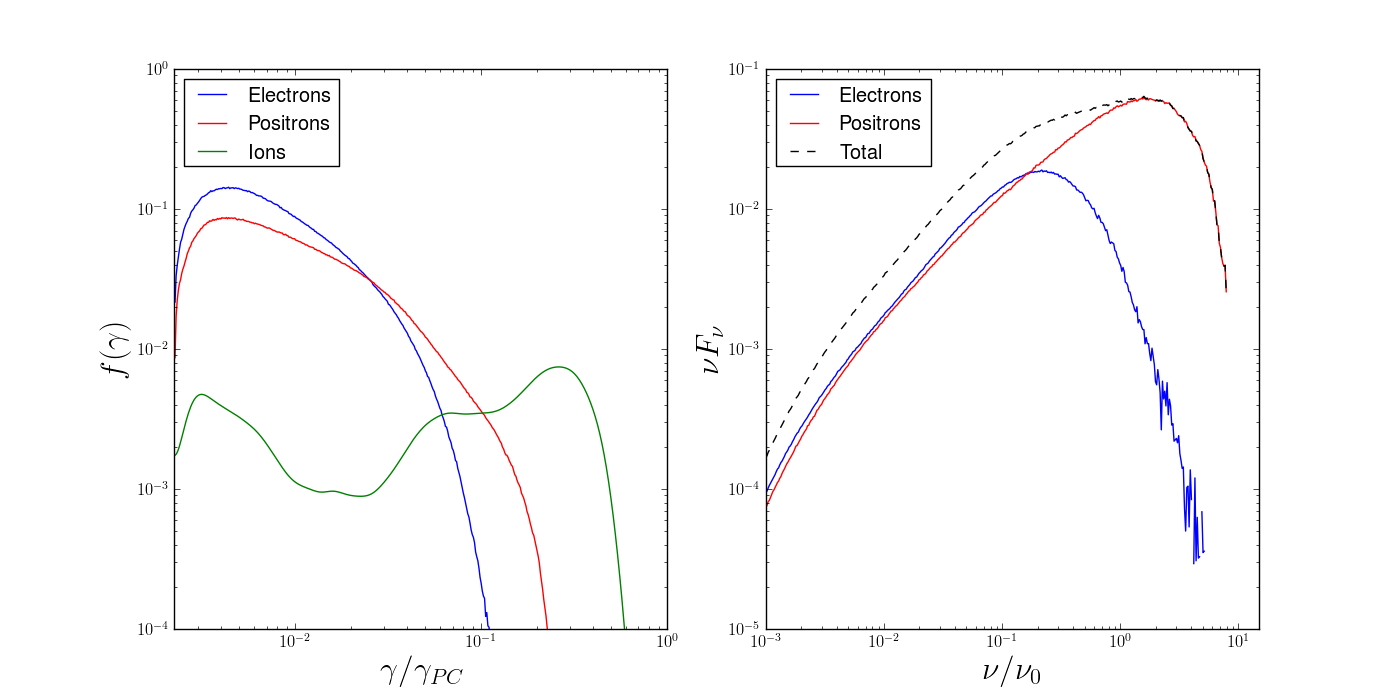} \\
     (a) $30^{\circ}$ magnetosphere \\
     \includegraphics[scale=0.4]{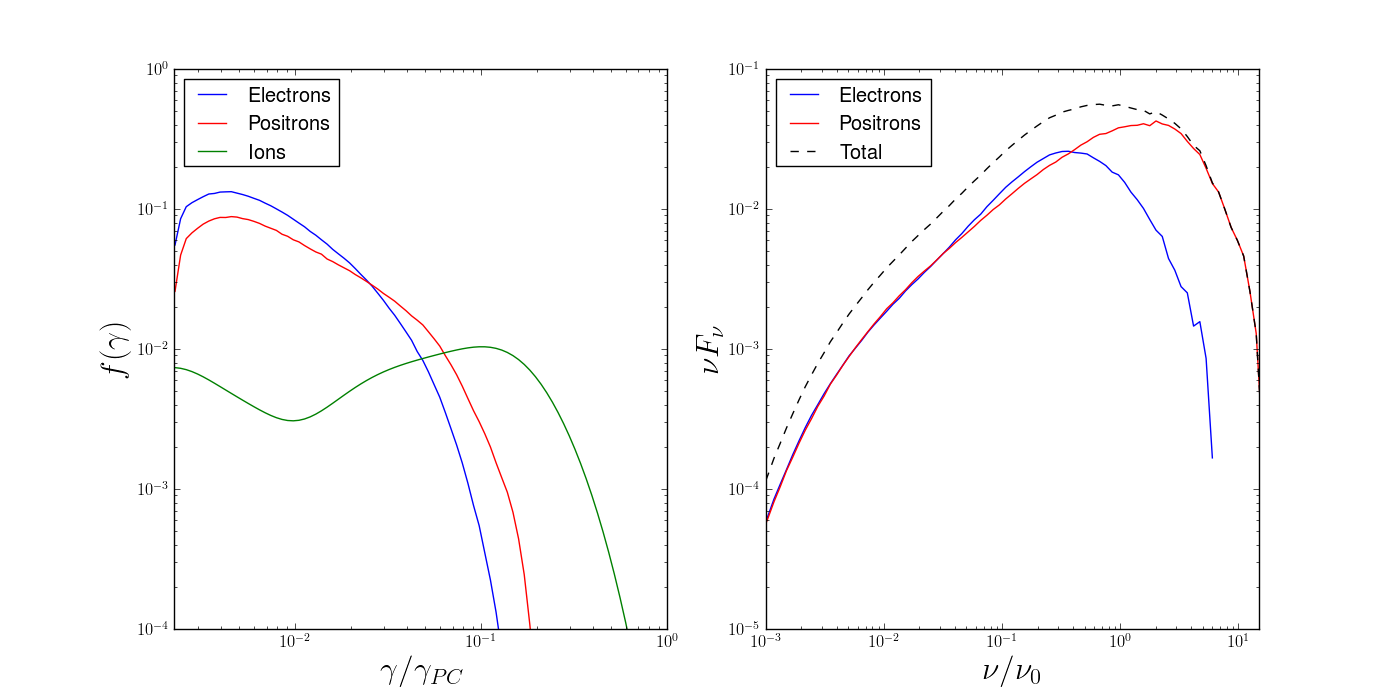} \\
     (b) $60^{\circ}$ magnetosphere \\
     \includegraphics[scale=0.4]{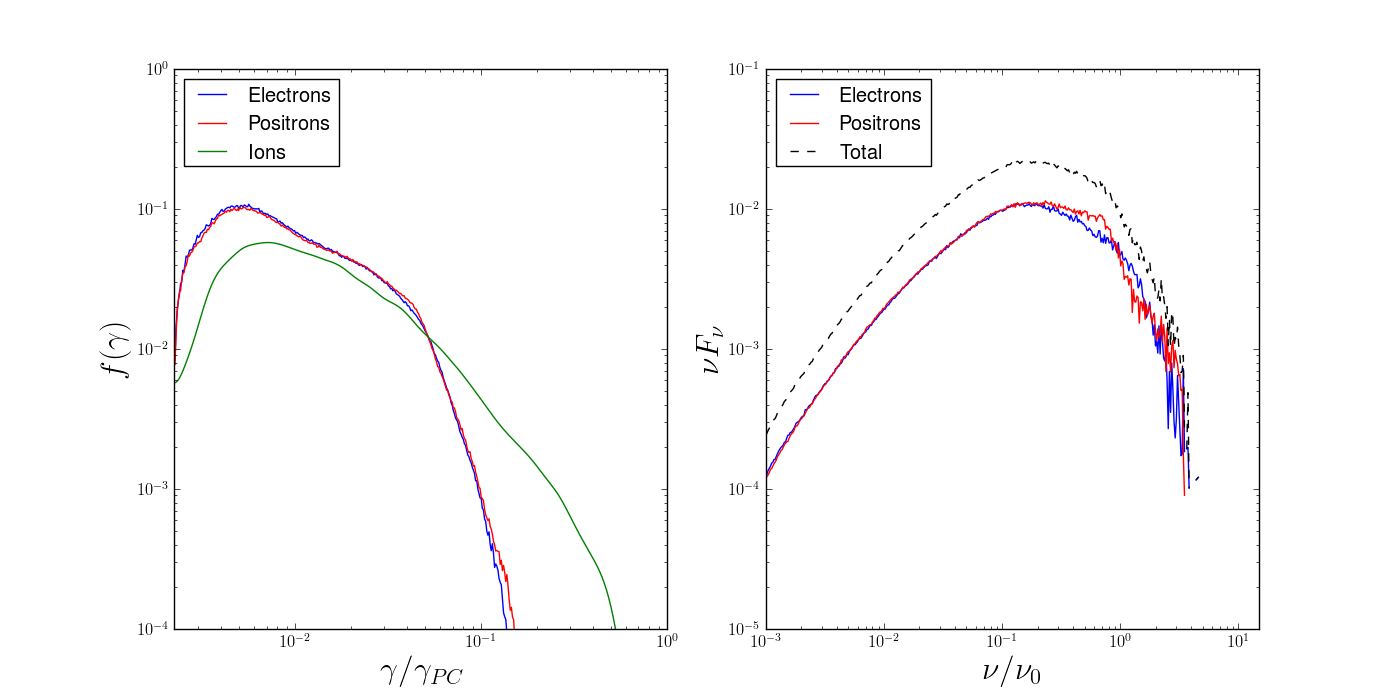} \\
     (c) $90^{\circ}$ magnetosphere \\
\end{tabular}
\caption{Particle and phase-averaged radiation spectrum of the current sheet in the magnetosphere. For particle spectrum, red, blue and green lines represent positrons, electrons and ions, respectively. For radiation spectrum, red, blue and green lines represent positron, electron and total emission, respectively. The results are shown for a range of inclination angles: (a) $30^{\circ}$, (b) $60^{\circ}$, (c) $90^{\circ}$.}
\label{fig:spect}
\end{figure*}

Earlier simulations of magnetospheres with pair production \citep{Chen14,Philippov15b} found the formation of a density gap above and below the current sheet. We find that if the photon mean free path becomes comparable to the light cylinder radius, this gap starts to be filled with plasma, produced in the collisions of photons emitted by energetic particles in strong current layers. However, we do not observe a subsequent discharge from the freshly produced pairs, so the density of pairs in this region remains significantly lower compared to the dense polar outflow and current layers. This is because the available voltage on these field lines is significantly smaller than the voltage in strong current sheets.

Pair production at high altitudes occurs through the two-photon pair production. We find that the region around the Y-point produces most of the simulation photons. As shown in Figure \ref{fig:photons30}, the region near the light cylinder also supports a counter-streaming photon distribution. We find that the backward propagating photons are emitted mostly by electrons, which were captured by the current sheet and reversed by the outward pointing reconnection electric field. The outgoing photons near the Y-point are mostly produced by the escaping positrons\footnote{We are considering the case $\vec{\Omega}\cdot\vec{B}>0$}, which come from the return layer or the wind. Counter-streaming photon flows are highly favorable for pair production. \citet{Lyubarsky96} was the first to estimate the pair production efficiency in the current sheet. This analysis can be improved by taking backward propagating photons into account.

We observe the equatorial current sheet to be unstable to both drift-kink and tearing instabilities. Plasmoids, which are produced by the tearing instability of the current sheet, are noticeable in three-dimensional current density distribution, as shown in Figure \ref{fig:particles}c. The drift-kink instability mostly disappears at high inclinations (see the difference in the current sheet shape in Figure \ref{fig:densities} and Figure \ref{fig:30ph}), since $\nabla \times {\bf B}$ is supported there mainly by the displacement current.

\subsection{Particle acceleration}
\label{sect:acceleration}

Near the star, particles are accelerated in regions of super-GJ current, $j_{\parallel}/j_{GJ}>1$, and in the return current layer, where $j_{\parallel}/j_{GJ}<0$ \citep{Beloborodov08, Timokhin13}. These regions are highlighted in the distribution of the magnetospheric current presented in Figure \ref{fig:discharge}a. As a consequence of efficient particle acceleration, $e^{\pm}$ discharge is ignited. We show particle phase-space distribution on field lines which carry super-GJ and return current in Figure \ref{fig:discharge}b and c, respectively. In both regions the accelerating voltage fluctuates around the value set by the pair production threshold. In the bulk of the polar cap, where $j_{\parallel}/j_{GJ}>1$, electric field accelerates electrons outwards and pushes positrons inward. As the voltage gets screened by freshly produced pairs, acceleration of the primary particles ceases, and the pair cloud escapes from the discharge zone. We also observe particles leaking from the cloud and propagating back to the star \citep{Timokhin13}. In the return current layer, accelerating electric field extracts ions from the stellar surface, which are unable to produce pairs. In this case, electrons which return back from previous discharge episodes or the Y-point, accelerate toward the star and initiate the discharge. Freshly produced positrons are accelerated outward, similar to ions, and start to produce pairs. We find that an analogous situation happens in the half of the polar cap of the orthogonal rotator where $\rho_{GJ}>0$ and ions are extracted. Thus, models with free particle escape are consistent with observations of radio interpulses, where both parts of the polar cap are observed to power coherent radio waves \citep{Beskin90}. Finally, in Figure \ref{fig:discharge}d we show the phase-space distribution on field lines which carry sub-GJ current, $0<j_{\parallel}/j_{GJ}<1$. Here, the accelerating electric field is screened by the multi-streaming charge-separated flow of electrons with low energies, and the disсharge is not ignited. As it is clearly seen in Figure \ref{fig:discharge}a, the non-stationary behavior of the polar discharge results in fluctuations of the current density and, thus, magnetic field components. As we noted in \citep{Philippov15b}, these fluctuations in the non-steady pair plasma flow could be the primary source for the observed coherent pulsar radio emission.

The most energetic particles in our simulations are ions extracted from the footpoints of the return current layer, see Figure \ref{fig:particles}a. In this Figure colored thick lines show particle trajectories in the corotating frame, and the color shows particle's Lorentz factor, with magenta color representing highest energies. Initial acceleration of ions happens at the base of the return current zone at the edge of the polar cap, where $e^{\pm}$ discharge operates\footnote{In the northern part of the polar cap in the northern hemisphere, the return current represents a narrow sheet, whose thickness is of the order of plasma skin depth. In the southern part of the polar cap, there is also a volume return current zone.}. However, they gain most of their energy at the Y-point and the current sheet beyond the light cylinder, where $E>B$ occurs. We find that the most energetic ions are accelerated close to the equatorial plane. Positrons which reach highest energies in our simulations are created in the $e^{\pm}$ discharge in the return current sheet, see Figure \ref{fig:particles}b. Most of them are injected into the Y-point and the current sheet at significantly smaller energies compared to ions and experience synchrotron cooling when they encounter strong magnetic field inhomogeneities inside the current sheet. However, maximum energy of positrons is still around $\sigma_{LC}$, significantly higher than the limit set by synchrotron radiation reaction. This is because deep inside the current sheet, where $B \to 0$ and particle acceleration happens, synchrotron losses are significantly reduced and do not limit the particle energy \citep{Cerutti2012,Kagan2016}. 

In Figure \ref{fig:particles}c we show different histories of electron energization. The major acceleration region of electrons is the polar cap, as is shown by the trajectory of electron with label A. Here, electrons are accelerated up to the pair formation threshold and initiate the pair discharge, which leads to the screening of the electric field. We find that many electrons in the magnetosphere of $60^{\circ}$ rotator do significant surfing around the current sheet surface and are later accelerated outwards (trajectory of a representative electron of this sort is labelled as B in Figure \ref{fig:particles}c). These electrons contribute to the flux of high-energy emission. Not all electrons in the current sheet of the oblique rotator are capable to escape from the magnetosphere. As in earlier studies, we observe electrons captured from the wind by the electric field in the current sheet outside the light cylinder, returning back to the star (see trajectory of the electron C in Figure \ref{fig:particles}c). We also observe some electrons that are produced in the decay of outgoing photons inside the current sheet and reversed by electric field back to the star (see trajectory of the electron D in Figure \ref{fig:particles}c). 

In Figure \ref{fig:spect} we show the particle spectrum in the current sheet beyond the light cylinder for different inclinations. The particles were selected from grid points where the magnitude of local current density is sufficiently high\footnote{Same criterion was used to plot the volume rendering of current density in Figure \ref{fig:particles}.}. Spectrum of electrons and positrons shows a clear power-law tail with index $\approx 1.5$, typical of relativistic magnetic reconnection \citep{Sironi14,Guo14, Werner16spec}. The spectrum extends from $\gamma_s \approx 2$, which corresponds to the injection energy of secondary pairs, until the limit determined by the magnetization parameter at the light cylinder
\begin{equation}
m_e c^2\sigma_{LC} \propto B_{LC} (c/\Omega)/ \lambda \propto m_e c^2\gamma_{PC} / \lambda.
\end{equation}
Our simulation with $L_{mfp}=0.5 R_{LC}$  reaches $\lambda \approx 10$ in the current sheet outside the light cylinder, with $\sigma_{LC}\approx 30$. For low inclination angles, the direction of the reconnecting electric field in the current sheet does not significantly depend on the distance from the Y-point and is mostly radial. This results in a significant excess of outflowing positrons at highest energies \citep{Cerutti15}. In current sheets of highly oblique rotators, the reconnecting electric field is mostly along $\theta$ direction. However, close to the Y-point electric field still has a significant radial component, which results in the excess of positrons at highest energies. The peak of the ion energy distribution decreases with increasing inclination angle, being in the range $0.1-0.3 \Phi_{PC}$.

\subsection{Pulsar wind}

\begin{figure*}
\centering
\includegraphics[trim=70px 0px 70px 0px, clip=true, scale=0.7]{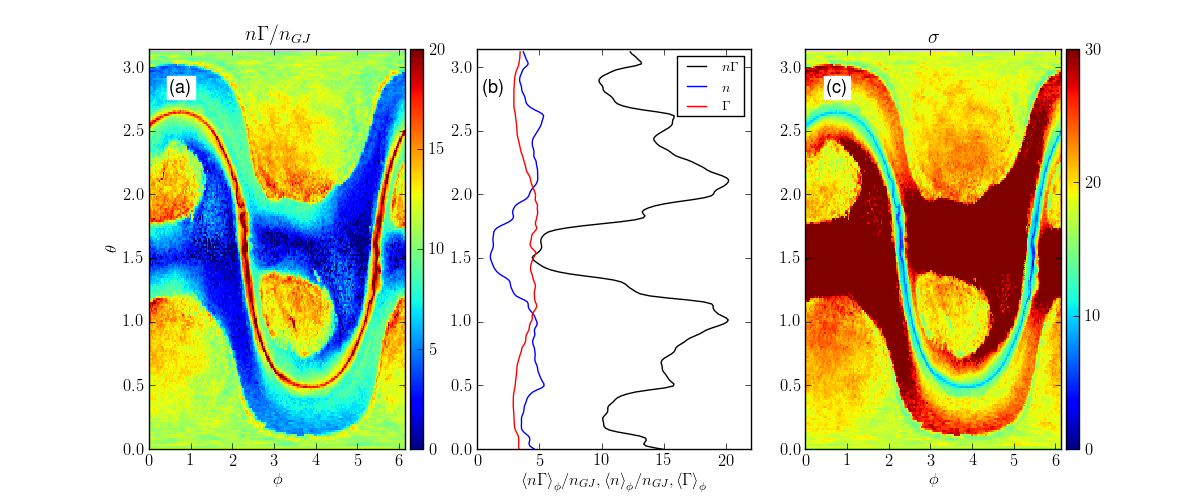}
\caption{Pulsar wind structure on a sphere of radius $3 R_{LC}$ in the magnetosphere of $60^{\circ}$ inclined rotator. (a): Plasma energy flux distribution. The distribution is highly non-uniform, because pair production does not operate on all field lines. (b): Plasma energy flux, density and bulk Lorentz factor averaged over azimuth. (c): Distribution of the magnetization parameter.}
\label{fig:pwind}
\end{figure*}

The structure of the pulsar wind beyond the light cylinder is highly non-uniform. In agreement with earlier conclusions of \citet{Tchekhovskoy16}, we find that the Poynting flux is strongly concentrated towards the equator, ${\langle S_r\rangle}_{\phi} \propto \sin^4 \theta$, for inclination angles $\alpha>40^{\circ}$. More interestingly, we find that the plasma energy flux and density distribution is also very non-uniform. In Figure \ref{fig:pwind}a we show the plasma energy flux, $m_e c^2\langle{n\gamma}\rangle$, on a sphere of radius $3 R_{LC}$. In Figure \ref{fig:pwind}b we present the plasma energy flux, density and bulk Lorentz factor averaged over azimuth\footnote{Since our solution is quasi-steady in the corotating frame, this coincides with the average over pulsar's rotation period}. Though the bulk Lorentz factor is highest at the equator due to heating in reconnection, the averaged energy flux is lowest there since the plasma density is significantly reduced. This can be understood as follows. First, as we discussed in \S\ref{sec:magnetosphere} and \S\ref{sect:acceleration}, there are regions on the polar cap which do not launch pair cascades and are sustained by a charge-separated outflow. Second, there is a density gap just above and below the current sheet, see discussion in Section \ref{sec:magnetosphere}. Though including finite mean free path of pair producing photons into account makes the gap filled with plasma, its density is still significantly lower compared to the dense polar outflows and equatorial current sheet.  

Since conditions at the base of the wind are now understood, the long-standing problem of whether the striped pattern of the pulsar wind survives until the termination shock \citep{Usov75,Lyubarsky01,Arons2012} can now be addressed. If the wind structure found in our simulations extends up to large distances, the non-uniformity of the pulsar wind at its base may have important implications for the particle acceleration at the termination shock. For example, our results imply that $\sigma$ parameter is highest at the equator, where particle energy flux is significantly reduced (see Figure \ref{fig:pwind}c). Since the gap above the current sheet is filled with plasma of density significantly lower compared to the polar outflow, the intermediate latitudes may support a high $\sigma$ region as well.

\section{Modeling of the high-energy emission}
\label{sec:highenergy}
\begin{figure*}
\centering

\includegraphics[trim=60px 8px 100px 40px, clip=true,width=\textwidth]{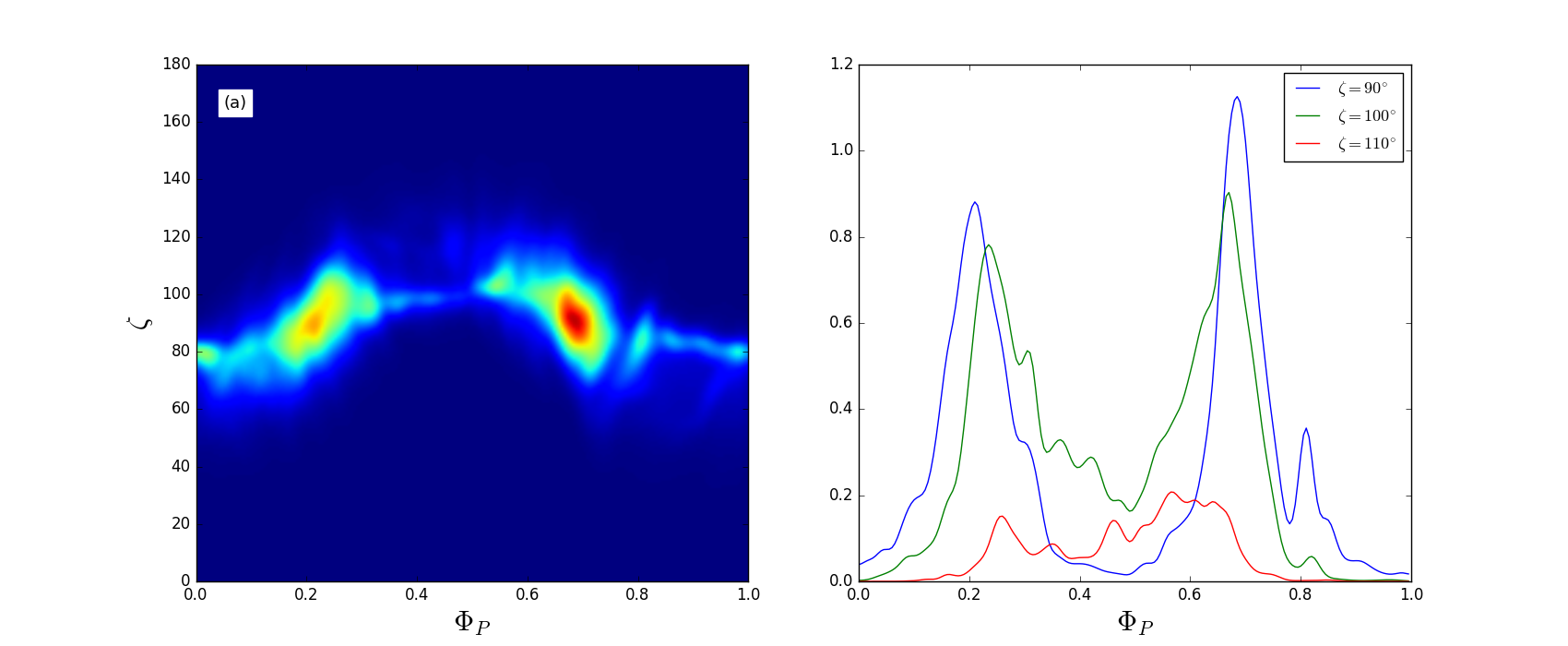}
\includegraphics[trim=60px 5px 100px 40px, clip=true,width=\textwidth]{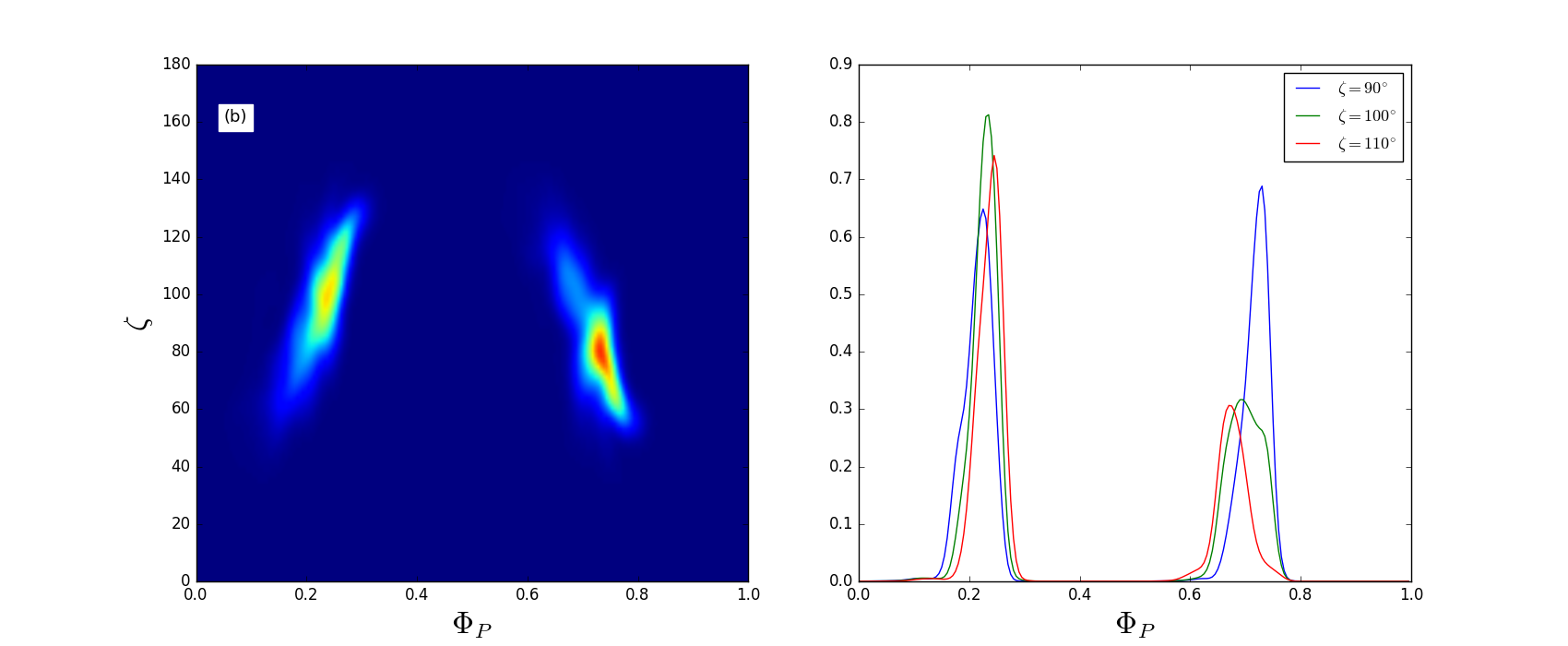}
\includegraphics[trim=80px 8px 140px 60px, clip=true,width=\textwidth]{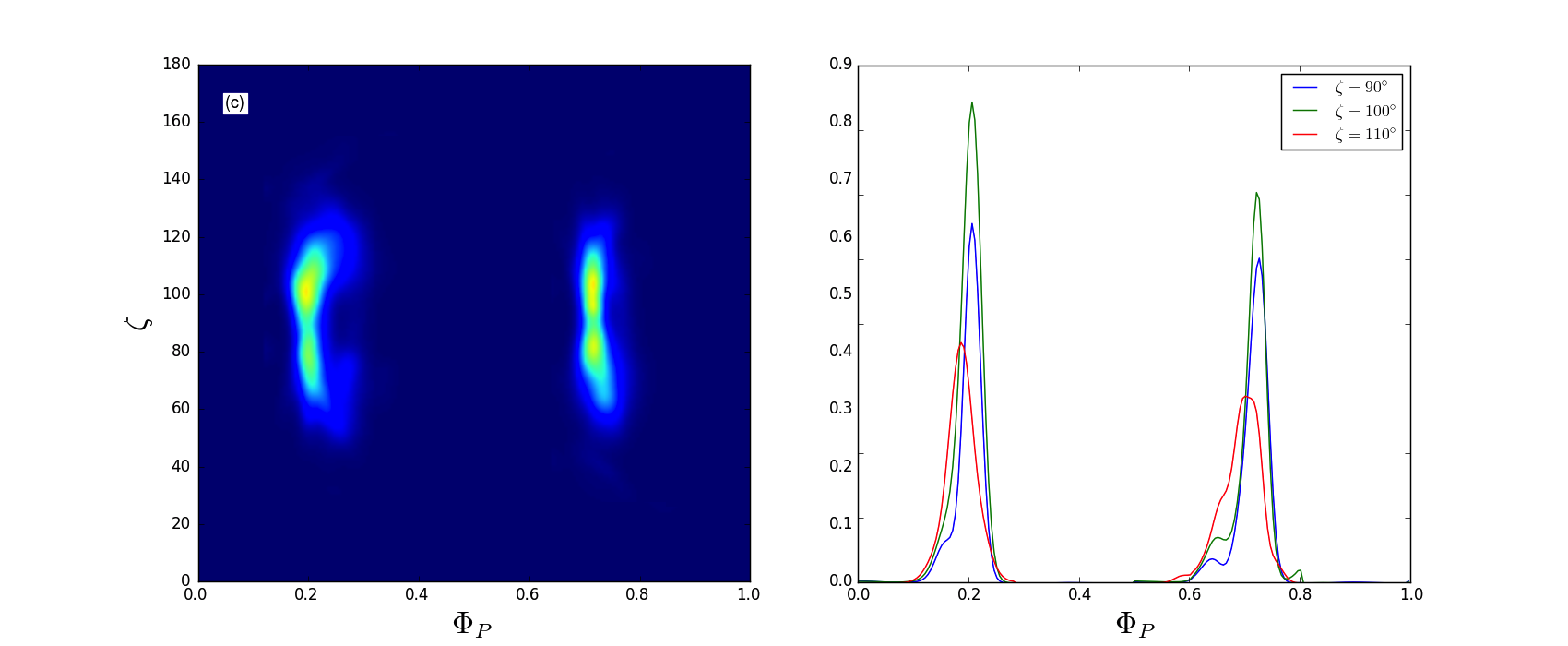}
\caption{Skymaps and light curves of high-energy emission. (a): $30^{\circ}$ inclined magnetosphere, (b): $60^{\circ}$ inclined magnetosphere, (c): $90^{\circ}$ inclined magnetosphere.}
\label{fig:LC}

\end{figure*}

The distribution of photons in Figure \ref{fig:densities}c shows that the high-energy emission is produced close to the Y-point and in the current sheet beyond the light cylinder. We find that the resulting light curves are similar to PIC simulations without pair production \citep{Cerutti16}, which suggests that the beaming of high-energy emission is mainly determined by the magnetospheric geometry rather than by the small-scale plasma processes. In Figure \ref{fig:LC} we show the skymaps of high-energy emission for inclination angles $30^{\circ},60^{\circ}$ and $90^{\circ}$, calculated by collecting the high-energy photons on a sphere of radius $2 R_{LC}$ over one rotation period. At low inclination angles, $\alpha<45^{\circ}$ the skymap shows a sinusoidal-like caustic, which traces the shape of the current sheet. The caustic is not uniformly bright, showing significant flux enhancement closer to the equator. In this geometry, we expect pulsars to produce mainly double-peaked gamma-ray emission profiles, most pronounced at viewing angles $\zeta\approx 90^{\circ}$. For intermediate observer viewing angles, $90^{\circ}<\zeta<90^{\circ}\pm\alpha$, double peaks are accompanied by significant bridge emission (see $\zeta=100^{\circ}$ in Figure \ref{fig:LC}a). For $\zeta\approx \pi/2\pm\alpha$ single-peaked profiles centered on $\Phi_P\approx 0.5$ are possible (see $\zeta=110^{\circ}$ in Figure \ref{fig:LC}a). As the inclination angle increases, the two bright equatorial parts become disconnected from each other, see Figures \ref{fig:LC}b and d. In this case, only the double-peaked profiles are expected. Due to the non-uniform brightness of the caustic the lightcurve is not symmetric. For $\zeta<90^{\circ}$ the second peak is brighter, while for $\zeta>90^{\circ}$ the first peak dominates. 

Synchrotron emission of the particle distribution in the current sheet shows a broad-band spectrum, rising in $\nu F_{\nu}$ at low energies, see Figure \ref{fig:spect}. The low-energy cutoff of the spectrum $\approx eB_{LC}\gamma^2_{sec}/m_e c$ is determined by the injection energy of secondary pairs $\gamma_{sec}\approx 2$. High energy positron excess results in flattening of the emission spectrum at the highest energies. Its peak is at approximately $\nu_0 = 0.1 e B_{LC} \sigma^2_{LC}/m_e c$, where $\sigma_{LC}$ takes into account the density of plasma, produced inside the sheet. In our simulations we find that the produced plasma multiplicity increases with the increased value of the magnetic field, $\lambda\propto B_{LC}$, so the magnetization parameter $\sigma_{LC}$ does not significantly change. Thus, the cutoff of the high-energy emission scales linearly with $B_{LC}$. While the highest energy part of the spectrum shows increasing $\nu F_{\nu}$ for low inclinations, $\alpha<40^{\circ}$, it is mostly flat or decreasing for larger inclinations. 

Everything else being equal, the radiative efficiency decreases with increasing inclination angle, which may partially explain the observed scatter in $L_{\gamma}$ vs $\dot E$ \citep{Abdo10}. The gamma-ray luminosity also depends on the efficiency of pair production around the light cylinder, which in our setup depends on the mean free path of photons, $L_{mfp}$, and on the strength of the magnetic field at the light cylinder. The maximum efficiency corresponds to large values of the mean free path, $L_{mfp} > 2R_{LC}$, in which case we reproduce the results of \citet{Cerutti16}, where pair production in the current sheet was neglected. In this regime only a very small number of photons is converted into pairs in our simulations. The radiative efficiency, $L_{\gamma}/\dot E$, varies between $\approx 10\%$ for the aligned rotator and $\approx 1\%$ for the orthogonal rotator, and scaling $L_{\gamma}\propto \dot E \sim B^2_{LC} R^2_{LC}$ holds. 

As the pair production in the sheet becomes efficient, the scaling $L_{\gamma}\propto \dot E$ breaks down, and radiative efficiency decreases for increasing $B_{LC}$. The physical reason can be understood as follows. As the pair density increases, the voltage drop available for the particle acceleration at X-points \citep{Bessho07,Sironi14,Werner16spec}, $\sim E_{rec} \lambda_p \sim \sigma_{LC}$, becomes limited\footnote{In the strong cooling regime particles are only accelerated at X-points. In plasmoids energetic particles quickly cool down, and further possible acceleration via Fermi process in island mergers \citep{Guo14,Sironi16} is not efficient.}. In the strong cooling regime significant part of particle's energy is radiated away, so we estimate the particle radiation power to be comparable to the acceleration rate due to the reconnection electric field, $\sim 0.1 e B_{LC} c$.  In this case the radiative power scales as $L_{\gamma} \sim e B_{LC}c n_{em} V_{em}$, where $n_{em}\sim n^{GJ}_{LC}$, $V_{em}\sim R^2_{LC}\lambda_p$ are number density of emitting particles and volume of the emitting region, respectively. Here we estimated the width of the current sheet to be of the order of the local plasma skin depth, and the radial extent of the emitting region to be comparable to the radius of the light cylinder. Since $0.1 e B_{LC}\lambda_p \sim \sigma_{LC}$, we obtain $L_{\gamma} \sim \sigma_{LC} n^{GJ}_{LC} R^2_{LC} \sim \sigma_{LC} \dot E^{1/2} \sim \dot E/\lambda$. As the pair multiplicity increases for higher $B_{LC}$, the radiative efficiency $L_{\gamma}/\dot E$ drops. Similarly, as the photon mean free path decreases, pair loading in the sheet becomes efficient, and the radiative efficiency drops. Although we expect the above calculation to be robust, we note that the accurate identification of scaling of the multiplicity parameter with magnetic field strength at the light cylinder requires accurate simulations of pair production in relativistic reconnection, including binary collisions of photons and realistic cross-sections. We will report this investigation elsewhere.

Our prescription for the pair formation mimics pair production in two-photon collisions in the outer magnetosphere. This process has the highest cross-section when the center-of-mass energy of colliding photons is around MeV. For example, this is satisfied for collisions of GeV and soft X-ray photons. The broad-band nature of the radiation spectrum from current sheets shows that pair-production can be self-sustained, e.g., soft X-ray photons, which are targets for further pair-producing collisions with high-energy GeV photons, are produced by secondary pairs in the sheet. This should be the case in young active pulsars, like the Crab, where the density of X-ray photons is sufficiently high. Older pulsars may sustain an active discharge with X-ray photons from the neutron star surface, similar to outer gap models.

\section{Discussion and conclusions}
\label{sec:discussion}

Our previous simulations \citep{Philippov15b} and subsequent analytical studies \citep{Gralla16,Belyaev16} showed that the effects of general relativity, in particular the frame-dragging, are essential in driving efficient pair production in aligned pulsars. Here we show that same effects help to increase the fraction of pair producing field lines in magnetospheres of oblique pulsars.

We show that plasma density and the energy flux in the pulsar wind is highly non-uniform. We find that if the mean free path of photons in the outer magnetosphere is comparable to the light cylinder, plasma density hole above the current sheet, identified previously by \citealt{Chen14, Philippov15b, Philippov15a}, is filled with pair plasma. However, since the voltage on these field lines is quite low, we do not observe an active $e^{\pm}$ discharge there. This is why the density of pair plasma in this region is much lower compared to the dense polar outflow and the current sheet. Given that conditions at the wind base are now understood, the $\sigma$ problem of the pulsar wind \citep{Coroniti90,Lyubarsky01} can be addressed from first principles. In particular, it's intriguing to investigate whether the non-uniform striped wind survives until the termination shock, or if it transforms to a vacuum-like wave \citep{Melatos1996} or transitions to turbulence \citep{Zrake2016}.

In agreement with earlier expectations \citep{Cerutti16}, particle acceleration in current sheets is not limited by synchrotron cooling, since energetic particles are focused deep inside the layer, where losses can be neglected. Our first-principles modeling of magnetospheric high-energy emission shows that pulsar gamma-rays are produced in strong current layers at the boundary of the closed field line region and beyond the light cylinder. As the obliquity of pulsar increases, the contribution of the separatrix current layer decreases, and pulsar gamma-ray emission is dominated by the synchrotron emission of particles accelerated by magnetic reconnection. Our calculations generally reproduce the double-peaked morphology of gamma-ray pulsars discovered by Fermi \citep{Abdo10}. We found that particle acceleration, cutoff in the spectrum of high-energy emission, and radiative efficiency are regulated by the efficiency of pair production near the Y-point and the current sheet. As the pair load of the sheet increases with higher magnetic field strength at the light cylinder, acceleration by reconnection electric field becomes limited, and radiative efficiency drops. This brings in the new fundamental problem in radiative reconnection research, e.g., how efficient is the pair production in reconnecting current sheets? We will approach this question with refined simulations in our future work.

We find that pulsars are efficient sources of energetic ions. For the geometry ${\bf{\Omega}} \cdot {\bf{B}}>0$ and inclination angles not very close to $90^{\circ}$, ions are extracted at the base of return current layer. Ions gain most of their energy at and beyond the Y-point. This is important for reaching extremely high energies, since ion acceleration in the return current layer at the boundary of closed field line region becomes limited due to curvature radiation losses \citep{Arons2012}. Deep inside the equatorial current sheet beyond the light cylinder, where ions are accelerated, this is not an issue. For the parameters of millisecond magnetars, this makes energetic ions interesting candidates for the ultra high-energy cosmic rays \citep{Arons2003}.

Previously, the concept of a ``weak pulsar" was introduced, e.g., a pulsar with sufficiently small magnetic field at the light cylinder to not drive efficient pair creation in the outer magnetosphere \citep{Chen14,Gruzinov13}. We performed axisymmetric simulations with order of magnitude higher polar cap voltages and multiplicities of the secondary plasma, and observed the return current being more and more active. We also observed the oscillations between the active force-free like solution and the disk-dome phase, which happens on a timescale of few rotational periods. This behavior can be interesting for modelling nulling in pulsars near the death line, which can not support large multiplicities in their pair cascades. The details of this study will be presented elsewhere.

We thank Jonathan Arons, Andrei Beloborodov, Vasily Beskin, Samuel Gralla, Beno{\^i}t Cerutti, Andrei Gruzinov, Alexander Tchekhovskoy, and Andrey Timokhin for fruitful discussions. This research was supported by the NASA Earth and Space Science Fellowship Program (grant NNX15AT50H to AP), Porter Ogden Jacobus Fellowship awarded by Princeton University to AP, NASA grant NNX15AM30G, Simons Foundation (grant 267233 to AS), and was facilitated by Max Planck/Princeton Center for Plasma Physics. The simulations presented in this article used computational resources supported by the PICSciE-OIT High Performance Computing Center and Visualization Laboratory and by NASA/Ames HEC Program (SMD-16-6663, SMD-16-7816).

\end{document}